# Mathematical model for hit phenomena as stochastic process of interactions of human interactions

Akira Ishii, Hisashi Arakaki, Naoya Matsuda, Sanae Umemura, Tamiko Urushidani, Naoya Yamagata and Narihiko Yoshda§

Department of Applied Mathematics and Physics, Tottori University
Koyama, Tottori 680-8552, Japan

§ Digital Hollywood University, Sotokanda, Chiyoda-ku, Tokyo 101-0021, Japan

E-mail: ishii@damp.tottori-u.ac.jp
       ishii.akira.t@gmail.com

# Mathematical model for hit phenomena as stochastic process of interactions of human interactions

**Akira Ishii, Hisashi Arakaki, Naoya Matsuda, Sanae Umemura, Tamiko Urushidani, Naoya Yamagata and Narihiko Yoshda§**

Department of Applied Mathematics and Physics, Tottori University
Koyama, Tottori 680-8552, Japan
§ Digital Hollywood University, Sotokanda, Chiyoda-ku, Tokyo 101-0021, Japan

E-mail: ishii@damp.tottori-u.ac.jp

**Abstract.** Mathematical model for hit phenomena in entertainments in the society is presented as stochastic process of interactions of human dynamics. The model use only the time distribution of advertisement budget as input and the words of mouth (WOM) as posting in the social network system is used as the data to compare with the calculated results. The unit of time is daily. The WOM distribution in time is found to be very close to the residue distribution in time. The calculations for Japanese motion picture market due to the mathematical model agree very well with the actual residue distribution in time.

## 1. Introduction

The human interaction in real society can be considered as many body theory of human. Especially after the population of the social network systems (SNS) like blogs, Twitter, Facebook, Google+ or other similar services in the world, the interactions between each accounts can be stocked as digital data. Though the SNS society is not equal to the real society, we can assume that the communication in SNS society is very similar to that in the real society. Thus, we can use huge amount of stock of digital data of human communication as observation data of the real society.[1-4] Using this observation, we can apply the method of statistical mechanics to social sciences. Since the word-of-mouth (WOM) has bee pointed out to be very significant for example in marketing science[5-8], such analysis and prediction of the digital WOM in the sense of statistical physics will be important today.

In this paper, as an applied field of the statistical mechanics of human dynamics, we focus our attention to the motion picture entertainment, because the logs of communication for each movie in SNS can be easily distinguished and the competition of markets between movies can be neglected because of the character of each movie; market of the Harry Potter, the Pirates of Carribian and the Avatar can be distinguished as different market, for example. Moreover, traditionally, the motion picture industries in Japan have daily data of revenue for each movies.

The theoretical treatment of motion picture business has long history in social sciences as the marketing science. The traditional way of the model to forecast motion picture revenue is to assume the following simple model,

$$R = ABCe^D \qquad (1)$$

where A, B, C and D are, for example, budget of advertisement of the movie, strength of the words of mouth (WOM), the impression of starring, etc. Then the formulae is "liniarized" as follows.

$$\log R = \alpha_1 \log A + \alpha_2 \log B + \alpha_3 \log C + \alpha_4 D \tag{2}$$

Using huge stock of market data, coefficients $\alpha_1, \alpha_2, \alpha_3 \alpha_4$ are determined using the accurate statistics[9-15]. However, before discussing the actual determined coefficients, we physicists have question on the model of eq. (1) itself. The form of eq. (1) itself should be considered deeply and should be derived. Moreover, the model of eq. (1) has no way to include dynamics of human interactions into account because of its too simplified idea. In the actual society including the SNS society, the communication between humans has some dynamics behavior, so that we should use more realistic model to consider the aggregation behavior of communications in the society.

The stochastic process was tried to be applied for the forecast of motion picture revenues also [16], but the approach is still incomplete and should be continued more accurately using the data of blogs, Twitter or Facebook posting.

The better approach from the point of view of physicists is the so-called Bass model which was presented as the simple model of aggregation behavior of WOM in 1969[17,18] The key concept of the Bass model can be considered as diffusion equation; diffusion of WOM in society. A lot of modified Bass models have been presented to analyze WOM for the motion pictures [19,20]. In the Bass model, we consider the number of adopted persons at the time t, R(t). The number of non-adapted persons is calculated as N-R(t) where N is the number of persons in the market. If advertisement affect the people to adapt the products, we can write the increasing rate of it as

$$\frac{dR(t)}{dt} = p(N - R(t)), \tag{3}$$

where p is the probability for non-adopted persons to adopt the product per unit time due to the advertisement. The people can be affected also due to WOM from the adapted person. Thus, if we consider only the WOM effect, we find

$$\frac{dR(t)}{dt} = q(N - R(t))R(t), \tag{4}$$

where q is the probability for non-adapted persons to adopt the product per unit time due to WOM from the adapted persons. Thus, combining both effect, we can write as follows,

$$\frac{dR(t)}{dt} = (N - R(t))(p + qR(t)). \tag{5}$$

This is the equation of the Bass model.

In the Bass model, the advertisement is included only as the factor p. The many modified Bass model include the decreasing per time of the advertisement effect using the exponential decay function as follows,

$$\frac{dR(t)}{dt} = (N - R(t))\left(pe^{-\alpha(t-t_0)} + qR(t)\right), \tag{6}$$

where, t0 is the time of the release day of the product we concern. However, the real marketing actions continue for several weeks before the release. The above modified Bass model does not include such advertisement effects before the release.

Moreover, the above Bass model and the modified Bass model do not include rumor effects in the real society that are not described using the person-to-person two body

From the above brief review of the previous studies, we find that the effect of the advertisement and the WOM is included incompletely and the rumor effect is not included. Therefore, from the point of view of statistical physics, we present in this paper the model to include the three effects; the advertisement effects, the WOM effects and the rumor effects. The presented model is applied to the motion picture business in the Japanese market and we compare our calculation with the reported revenue and the observed number of posting blogs for each films.

## 2. Model
*2.1 Purchase intention*

We start the modeling from the viewpoint of the individual consumer level. We define the purchase intention of the individual consumer labeled i at time t as $I_i(t)$. We assume that the number of product adopted till the time t can be written as

$$Y(t) = p \int_{t_0}^{t} \sum_{i=1}^{N} I_i(t) dt \quad , \tag{7}$$

where N is the maximum number of adopted persons, p is the price of the product and $t_0$ is the release day of the product we concern. Thus, our problem is to define the equation of the purchase intention of each consumer $I_i(t)$. We consider the modeling of the effect of the advertisement, the WOM and the rumor for the purchase intention in the following subsections.

*2.2 Advertisement effect*

The advertisement effects due to mass media like TV, newspaper, magazine, web, the Facebook or the Twitter is modeled as an external force for the equation of the purchase intention of the individual consumer.

$$\frac{dI_i(t)}{dt} = c_i A(t) \quad , \tag{8}$$

where A(t) is the time distribution of the effective advertisement effect per each time and the coefficient $c_i$ describe the impression of the advertisement for the consumer i. The external force A(t) can be considered as trends of the world or political pressure to the market. In the application to the motion picture business in Japanese market, we input the real daily advertisement budget used by the largest advertisement office, Dentsu Inc. in Japan.

*2.3 WOM effect*

Usually, the film success is spread by word of mouth (WOM). Such WOM is sometimes very significant effect to the hit of the movie. Thus, such WOM effect should be included in our theory.

The WOM effect should be distinguished into the two types; WOM direct from friends and indirect as rumor. We name that the WOM effect from friends is *direct communication,* because customers obtain information directly from their friends. Usually in previous marketing theories based on the Bass model[17-20], only the communications from the adaptor to non-adaptor are take into account. Here, in this paper, we include also the communication between non-adaptors. It is very sisnificant for movie entertainment especially before the open of the movie. Let consider that the person "i" hear the information form the person "j". The probability per unit time to effect the information to the purchase intention to the person "i" can be described as $D_{ij} I_j(t)$ where $I_j(t)$ is the purchase intention of the person "j" and $D_{ij}$ is the coefficient of the direct communication. The schemaitic image of the direct communication of the persons i and j is shown in the figure 1. Thus, we can write the effect of the direct communication as follows,

$$\sum_{j=1}^{N} D_{ij} I_j(t) \tag{9}$$

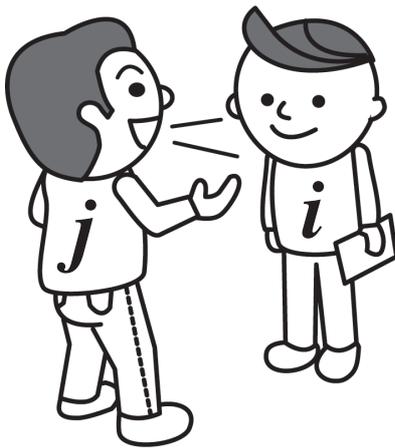

**Figure 1**
Schematic illustration of direct communication betwee the person i and the j.

Rumor effect is named in this paper as *indirect communication* where the person hears the rumor from chats on a street, chats from the next table in restaurant, chats in the trains or finds the rumor in an Internet blogs and Twitter. The situation is illustrated in figure 2a where many conversations are done in the street of a city. To construct the theory using mathematics, we focus to one person who listen a chat around him/her. Let consider that the person "i" hear the chat between the person "j" and the person "k", the strength of the effect of the chat can be described as $D_{jk}I_j(t)I_k(t)$. The probability per unit time to affect the chat to the purchase intention of the person "i" is defined as $Q_{ijk}D_{jk}I_j(t)I_k(t)$ where $Q_{ijk}$ is the coefficient. Thus, the indirect communication coefficient can be defined as $P_{ijk} = Q_{ijk}D_{jk}$. The situation we imaged is shown in figure 2b.

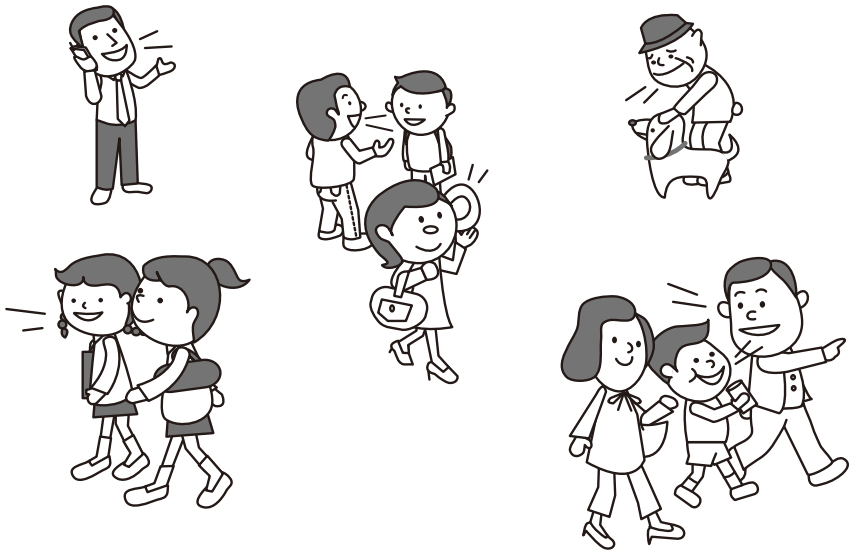

**Figure 2a**
The schematic illustration of communication in a city.

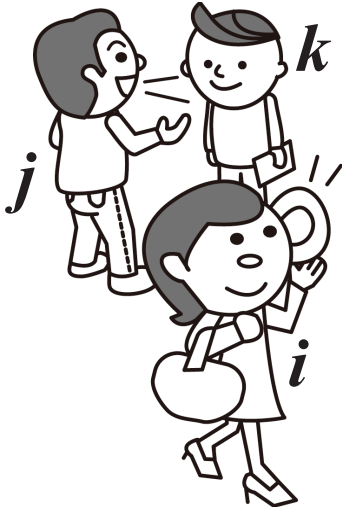

**Figure 2b**
The schematic illustration of the image of indirect communication in this paper. The person i listen a chat of the person j and k.

Therefore, the direct communication is the two-body interaction and the indirect communication is the three-body interaction. Thus, our theory for hit phenomena can be described as the equation of the purchase intention of a person i having the two-body interaction and the three-body interaction terms.

*2.4 Decline of audience*
For moving picture business, a person watches a certain movie only once. It means that the potential number of audience of box decreases monotonically after the opening day. In figure 3, we show the typical decline of cinema box office for Japanese movie market. It should be commented that the DVD disk business for a certain movie begin to supply at several monthes after the opening day of cinema box in Japanese entertainment market. Thus, the days shown in fig.3, no one can buy DVD or online movie in Japan.

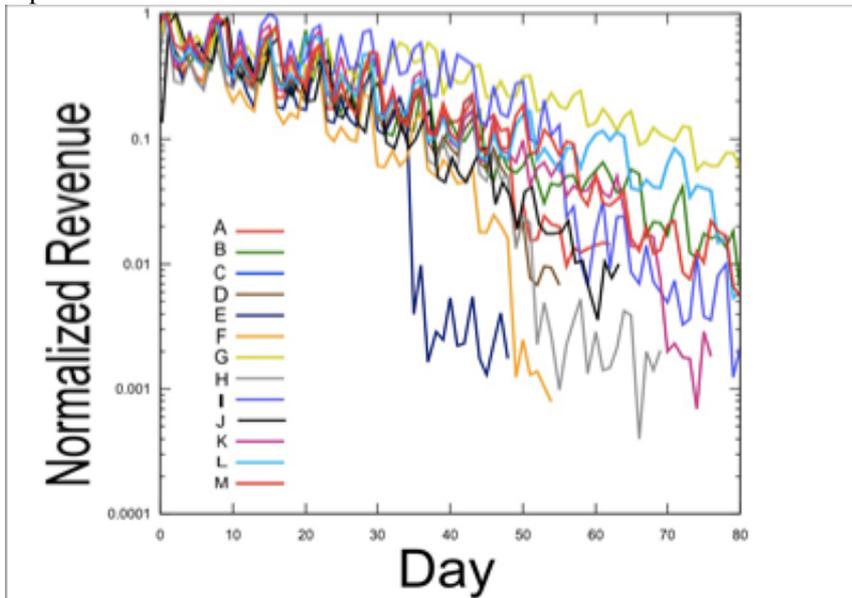

**Figure 3**
Revenues of the movies are shown as functions of date. Revenue data is the value in Japanese market. The movie titles are as follows. A:Death Note, B:Death Note the last name, C:zoku Always, D: Close zero, E: Closed Note, F:Biohazard 3, G:Bushi no Ichibun, H:Ratatouille,I:fra-girl, J:Dainihonjin, K:Maiko haaaan!!, L:Koizora, M:HERO.

We found in fig.3 that the revenues decrease almost exponentially. This evidence is very natural, because the number of audience decrease monotonically due to the effect that person who watched the movie does movies in fig.3 seems to be similar. The decay factor of the exponential decay is nearly 0.06 per day. It agrees well with the empirical rule of the Japanese movie market that the number of audience decreases roughly 6 percents.

This exponential decrease can be explained easily by using a simple mathematical model. First, we define that the number of potential audience to be $N_0$ and the number of integrated audience at the time t to be $N(t)$. Thus, the number of people who have interest on the movie is $N_0 - N(t)$. When we assume that the probability to watch the movie per one day is a, we obtain the equation to describe the number of audience to be

$$\frac{dN(t)}{dt} = a(N_0 - N(t)) \tag{10}$$

The solution of the equation with $N(0)=0$ at $t=0$ is

$$N(t) = N_0 (1 - e^{-at}) \tag{11}$$

It is clear that the result can be explain roughly the exponential decay of the audience shown in fig.3. The purchase-intention of a person i, $I_i(t)$ can be considered also to decay in same manner.

*2.5 Equation of purchase-intention for hit phenomena*

According to the above consideration, we write down the equation of purchase-intention in individual-level as the following manner,

$$\frac{dI_i(t)}{dt} = -aI_i(t) + \sum_j d_{ij} I_j(t) + \sum_j \sum_k h_{ijk} d_{jk} I_j(t) I_k(t) + f_i(t) \tag{12}$$

where $d_{ij}$, $h_{ijk}$ and $f_i(t)$ are the coefficient of the direct communication, the coefficient of the indirect communication and the random effect for the person i. We consider the above equation for every consumer so that $i=1,.. N_p$.

Taking the effect of the direct communication, the indirect communication and the decline of audience into account, we obtain the above equation for the mathematical model for hit phenomena. The advertisement and publisity effect for each person can be described as the random effect $f_i(t)$.

The equation (12) is the equation for all individual persons, but it is not convenient for analysis. Thus, we consider here the ensemble average of the purchase-intention of the individual persons as follows,

$$\langle I(t) \rangle = \frac{1}{N} \sum_i I_i(t) \tag{13}$$

Taking the ensemble average of eq.(12), we obtain for the left-hand side,

$$\left\langle \frac{dI_i(t)}{dt} \right\rangle = \frac{1}{N} \sum_i \frac{dI_i(t)}{dt} = \frac{d}{dt}\left(\frac{1}{N} \sum_i I_i(t)\right) = \frac{d\langle I \rangle}{dt} \tag{14}$$

For the right-hand side, the ensemble average of the first, the second and the third is as follows,

$$\langle -aI_i \rangle = -a \frac{1}{N} \sum_i I_i(t) = -a \langle I(t) \rangle \tag{15}$$

$$\left\langle \sum_j d_{ij} I_j(t) \right\rangle = \left\langle \sum_j d I_j(t) \right\rangle = \frac{1}{N} \sum_i \sum_j d I_j(t) = \sum_i d \frac{1}{N} \sum_j I_j(t) = Nd \langle I(t) \rangle \tag{16}$$

$$\left\langle \sum_j \sum_k p_{ijk} I_j(t) I_k(t) \right\rangle = \left\langle p \sum_j \sum_k I_j(t) I_k(t) \right\rangle$$

$$= \frac{1}{N} \sum_i p \sum_j \sum_k I_j(t) I_k(t)$$

$$= \sum_i p \frac{1}{N} \sum_j \sum_k I_j(t) I_k(t)$$

$$= Np \sum_i \frac{1}{N} \sum_j I_j(t) \frac{1}{N} \sum_k I_k(t)$$

$$= N^2 p \langle I(t) \rangle^2 \tag{17}$$

where we assume here that the coefficiet of the direct and the indirect communications can be approximated to be

$$d_{ij} \cong d$$

$$h_{ijk} d_{jk} = p_{ijk} \cong p$$

under the ensemble average.

For the forth term, the random effect term, we consider that the random effect can be divided into two parts; the collective-effect and the individual effect. I write

$$f_i(t) = \langle f(t) \rangle + \Delta f_i(t) \tag{18}$$

$$\langle f_i(t) \rangle = \frac{1}{N} \sum_i f_i(t) = \langle f(t) \rangle \tag{19}$$

where $\Delta f_i(t)$ means the deviation of the individual external effects from the collective effect, $\langle f(t) \rangle$. Thus, we consider here that the collective external effect term corresponds to the advertisement and the publicity to persons in the society. The deviation term $\Delta f_i(t)$ corresponds to the deviation effect from the collective advertisement and the publicity effect for the individual persons so that we can assume to be

$$\langle \Delta f_i(t) \rangle = \frac{1}{N} \sum_i \Delta f_i(t) = 0 \tag{20}$$

Therefore, we obtain the equation for the ensemble-averaged purchase-intention in the following manner,

$$\frac{d\langle I(t) \rangle}{dt} = -a \langle I(t) \rangle + D \langle I(t) \rangle + P \langle I(t) \rangle^2 + \langle f(t) \rangle \tag{21}$$

where

$$Nd = D$$

$$N^2 p = P \tag{22}$$

The equation (21) can be applied to the purchase-intention in the real market. The equation (21) is the equation we assumed in our previous works without derivation [21-25]. In this paper, we apply this equation to the motion picture business.

## 3. Observed data in the market

For the application of eq. (21) to real markets of motion pictures, we observe some market data as inputs and the observation compared with our calculation. The market data we use here are daily advertisement cost, daily revenue and daily number of posting of blogs in the Internet.

*3.1 Daily advertisement data*

we show the advertising cost and the related GRP (Gross Rating Point) for the movie "Da Vinci Code" in Japanese market. GRP is a term used in advertising to measure the size of an audience reached by a specific media vehicle or schedule [26]. It is just the product of the percentage of the target audience reached by an advertisement times the frequency they see it in a given campaign of the movie. Both the daily advertising costs and the GRP for each movies in Japanese market are obtained from Dentsu Inc. which is the biggest advertisement agency in Japan. In Japan, the most of all advertisement on every TV except NHK (National TV station) is handled by Dentsu, so that the whole campaign of every each movies at all Japanese prefectures are organized totally by Dentsu, so that the advertisement cost data from Dentsu is exactly equal to the advertisement costs of the movie in the Japanese market. The value of the daily advertising costs is not including the discount for major movie production agency in Japan, but the discount rate is same during the campaign of each movie.

We found clearly in fig.4 that the advertising costs and GRP are distributed for 2 months even after the open of the movie. This feature can be found for every movie campaign in Japanese market. Thus, not only the total amount of advertising cost but the time distribution of the costs are very important for the success of the advertising campaign of motion picture.

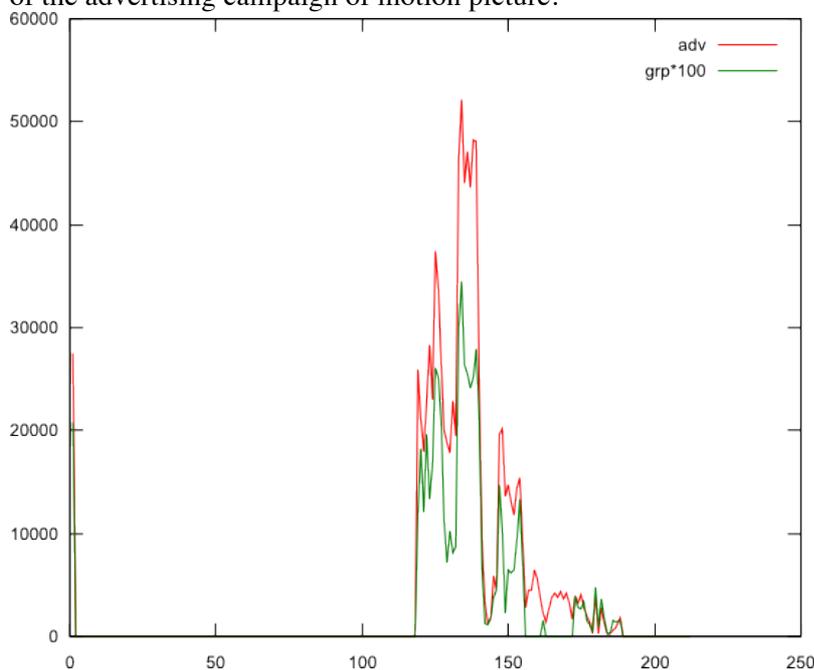

**Figure 4**
The advertising cost and the corresponding GRP for the movie Da Vinci Code in Japanese market. The open day is 140 of the horizontal cooridnate.

*3.2 Daily audience*
Number of daily audience for each movie is very important for the mathematical study of the motion picture campaign. In Japan, the sales of DVD disk or online supply of movies are forbidden till several months after the open of the movie in movie theaters. Thus, the number of audience we obtain is exact number of audience who watch the movie. In these duration, no one can watch movies without going to movie theaters. We obtain the data of daily audience from Box Office Japan (Kogyo Tsushinsha, Tokyo).

*3.3 Daily blog postings*
Daily posts of blogs for a certain movie is very important signal to measure the movement of purchase-intention of persons in the society. We measure daily data of the number of posts for the movies using the site, Kizasi in Japan which is the service of observing blog postings. Our measurement of the number of blog posts for several movies in Japan are shown in the following figures 5 - 9. In these figures, the curve of the daily blog posts are multiplied by a certain constant to normalize the value to be similar to the revenue value.

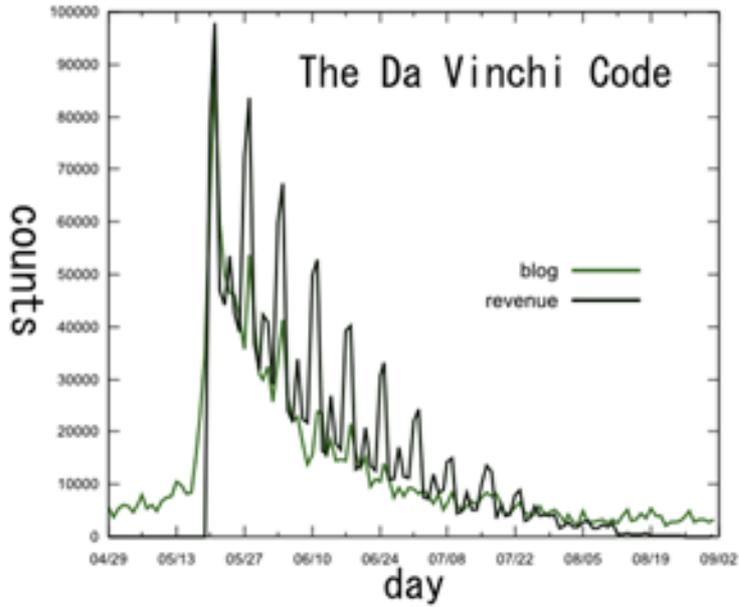

**Figure 5**
Daily revenue and posts of blog for Da Vinci Code. The black line is the revenue and the green line is the post of blog.

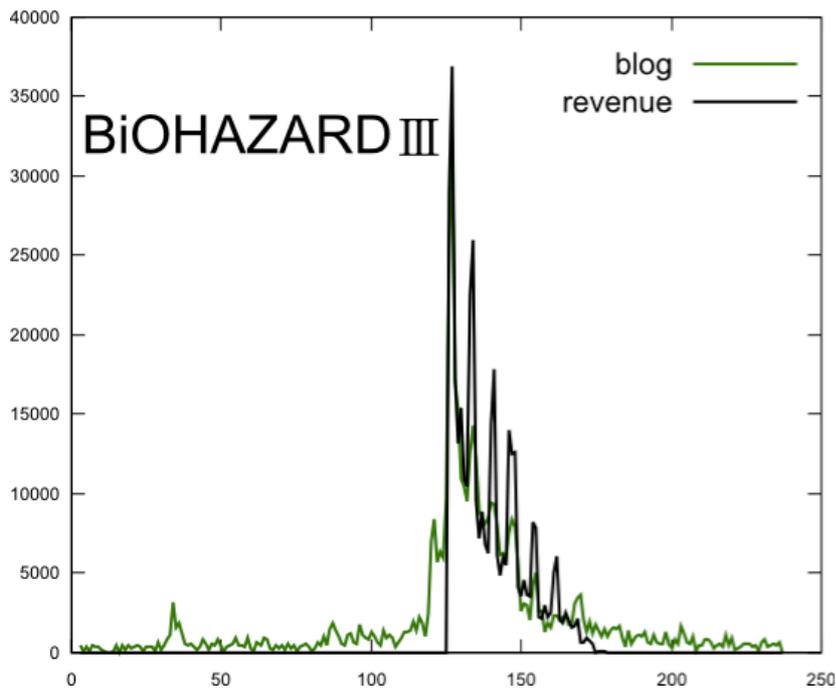

**Figure 6**
Daily revenue and posts of blog for Biohazard III. The black line is the revenue and the green line is the post of blog.

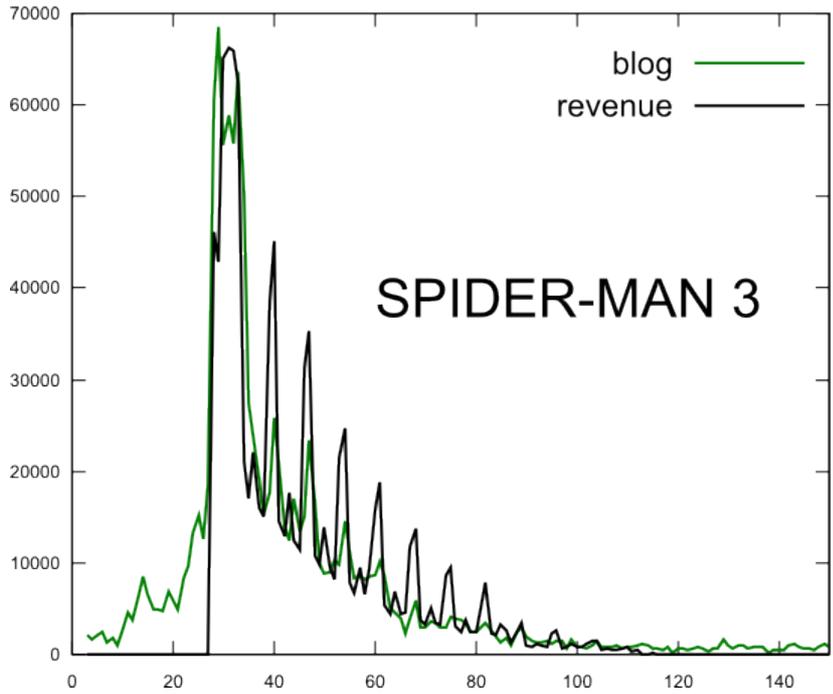

**Figure 7**
Daily revenue and posts of blog for Spider-man 3. The black line is the revenue and the green line is the post of blog.

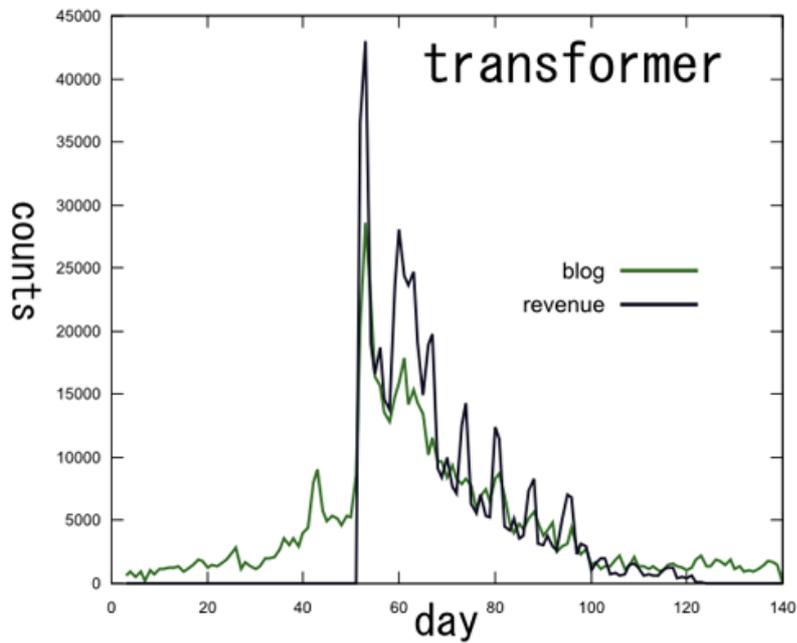

**Figure 8**
Daily revenue and posts of blog for Transformer. The black line is the revenue and the green line is the post of blog.

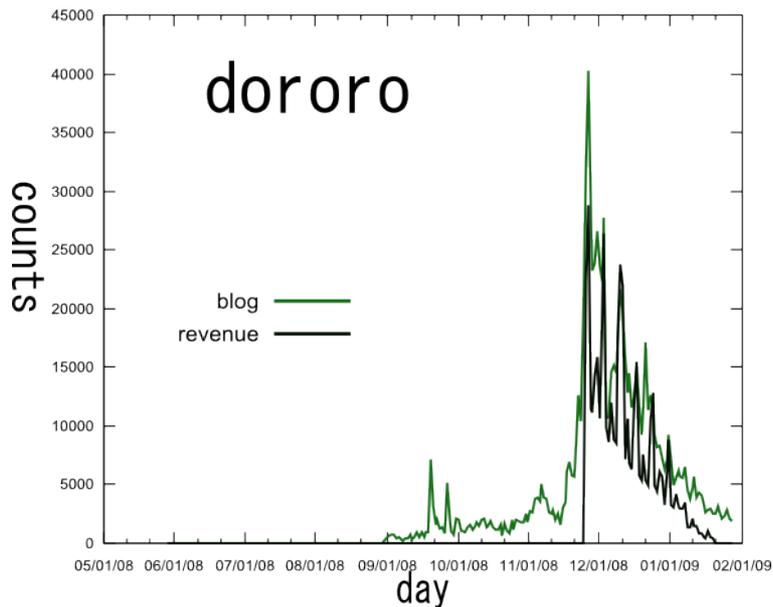

**Figure 9**
Daily revenue and posts of blog for Dororo(JPN). The black line is the revenue and the green line is the post of blog.

From figs.5-9, we found that the curves of the daily blog posts for each movies are very similar to the curve of the daily revenue. We observe for 25 movies in Japanase market and found the similar feature for every 25 movies. The observation means that the ratio of posting blogs for each person is almost constant during the duration of each movie so that the daily number of blog posting is very similar to the daily revenue.

Such blogs for each films are distinguished into positive, negative and neutral opinion. The positive opinion means he/she wants to watch the film or judge the watched film to be positive. According to the recent study for the Japanese motion picture blogs[27], more than half of the blogs shows positive opinion. Moreover, they found that the ratio of positive, negative and neutral is almost constant during the duration of the movie opening. Thus, the observed blog posting counts can be considered to be proportional to the counts of positive blog postings.

According to the observation, we propose to use the daily number of blog posts as the daily "quasi-revenue". The quasi-revenue is very useful for analysis, because the quasi-revenue can be defined even before the open of the movie. We can observe the increase of the expectation of persons in the soceity for a movie.

## 4. Calculation

We calculate the daily purchase-intention using eq.(21) with the daily advertisement cost as input data for $<f(t)>$. The parameters in eq.(21) are adjusted to fit the data with the observed quasi-revenue for each movies.

*4.1 Calculation in detail*

For the calculation for each movies, we should derive the actual formulation for the calculation where the adopted person and the non-adapted person are distinguished. For the direct communication term of non-adaptor to non-adaptor interaction in eq.(21), we can write as follows,

$$D\langle I(t)\rangle = \frac{1}{N}\sum_i \sum_j dI_j(t)$$
$$= \frac{1}{N}d\sum_i N\frac{1}{N}\sum_j I_j(t)$$
$$= \frac{N_p - N(t)}{N_p}(N_p - N(t))\frac{1}{N_p - N(t)}\sum_j d^{nn}I_j(t)$$
$$= \frac{N_p - N(t)}{N_p}(N_p - N(t))d^{nn}I(t)$$

(23)

where

$$N(t) = N_p \int_0^t \langle I(\tau)\rangle d\tau$$

(24)

The direct communication term for adaptor to non-adaptor is

$$D\langle I(t)\rangle \Rightarrow \frac{N(t)}{N_p}(N_p - N(t))d^{ny}I$$

(25)

Similarly, we obtain the indirect communication term due to the communication between the non-adapted persons at the time t,

$$P\langle I(t)\rangle^2 = p\frac{1}{N}\sum_i \sum_j \sum_k I_j(t)I_k(t)$$
$$= p\frac{1}{N}\sum_i N\frac{1}{N}\sum_j N\frac{1}{N}\sum_k I_j(t)I_k(t)$$
$$= \left(\frac{N_p - N(t)}{N_p}\right)^3 N_p^2 p^{nn}I^2$$
$$= \frac{(N_p - N(t))^3}{N_p}p^{nn}I^2$$

(26)

where $p^{nn}$ is the factor of the indirect communication between the non-adapted persons at the time t. For the indirect communication, we obtain more two terms corresponding to the indirect communication due to the communication between adaptors and that between adaptor and non-adaptor as follows,

$$\frac{(N(t))^2(N_p - N(t))}{N_p}p^{yy}I^2$$
$$+\frac{N(t)(N_p - N(t))^2}{N_p}p^{ny}I^2$$

(27)

where $p^{yy}$ is the factor of the indirect communication between adaptors and $p^{ny}$ is the factor of the indirect communication due to the communication between adaptor and non-adaptor at the time t.

Finally, we obtain the equation of the purchase-intention for the actual calculation as follows,

$$\frac{d\langle I(t)\rangle}{dt} = -a\langle I(t)\rangle + \langle f(t)\rangle$$

$$+ \frac{(N_p - N(t))^2}{N_p} d^{nn}\langle I\rangle + \frac{N(t)}{N_p}(N_p - N(t))d^{ny}\langle I\rangle$$

$$+ \frac{(N_p - N(t))^3}{N_p} p^{nn}\langle I\rangle^2 + \frac{(N(t))^2(N_p - N(t))}{N_p} p^{yy}\langle I\rangle^2$$

$$+ \frac{N(t)(N_p - N(t))^2}{N_p} p^{ny}\langle I\rangle^2$$

(28)

For the duration before opening of the movie, there are no adapted persons. Furthermore, the total number of adapted person is zero. Thus, for the duration before opening, the equation (28) is reduced to the following form.

$$\frac{d\langle I(t)\rangle}{dt} = -a\langle I(t)\rangle + \langle f(t)\rangle + N_p d^{nn}\langle I\rangle + N_p^2 p^{nn}\langle I\rangle^2$$

(29)

The equation (28) with (24) is the nonlinear integro-differential equation. However, since the handling data is daily, the time difference is one day. We can solve the equation numerically as a difference equation.

*4.2 Reliable factor*

For the purpose of the reliability, we introduce here the so-called "R-factor" (reliable factor) well-known in the field of the low energy electron diffraction (LEED) experiment [28]. In the LEED experiment, the experimentally observed curve of current vs. voltage is compared with the corresponding theoretical curve using the R-factor.

For our purpose, we define the R-factor for our purpose as follows,

$$R = \frac{\sum_i (f(i) - g(i))^2}{\sum_i (f(i)^2 - g(i)^2)}$$

(30)

where the function f(i) and g(i) is defined in figure 10. The smaller R, the function f and g show that better matches. We use this R-factor as guide to get best adjustment of our parameters for each movies.

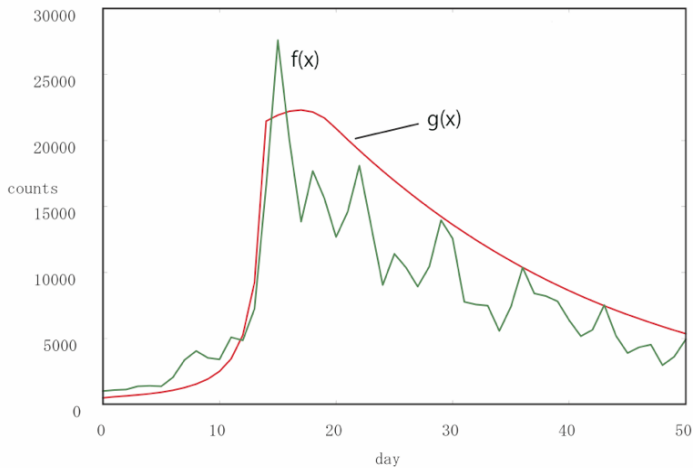

**Figure 10**
Reliable factor *R* is a guide to adjust the tow curves, *f(x)* and *g(x)*. R=0 as *g(x)* equal to *f(x)*.

*4.3 Results*

We perform calculations using eq.(29) for many movies movies in Japanese market. We show several results in fig.11-fig.15. The solid line shows our calculation with the real daily advertisement costs as input data of <f(t)> of eq.(29). In the calculation, we choose the parameters The parameters in the calculation are chosen to minimize the R-factor. The R-factor for each calculations are shown in the figure caption of fig.11-15. The results are compared with the number of the blog posting as the quasi-revenue shown as red histogram in the figures. We found that the agreement of the calculation with the quasi-revenue (blog) is very well.

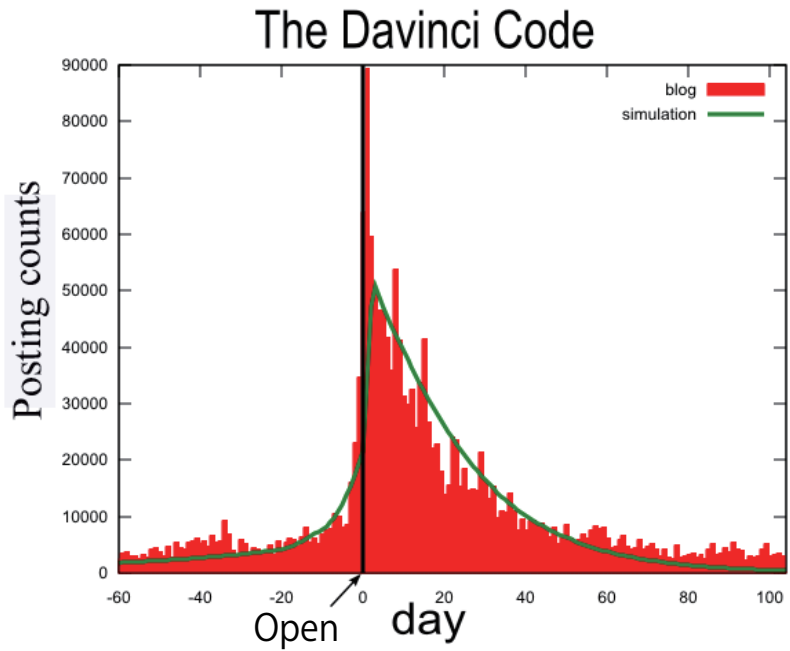

**Figure 11**
Calculation (solid line) and blog posting (histogram) for The Da Vinci Code. R=0.079

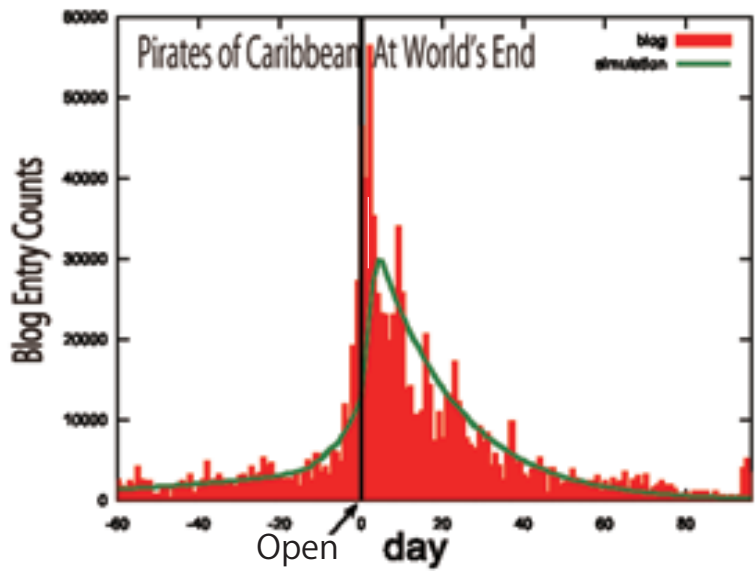

**Figure 12**
Calculation (solid line) and blog posting (histogram) for Pirates of Caribbean At World's End. R=0.135

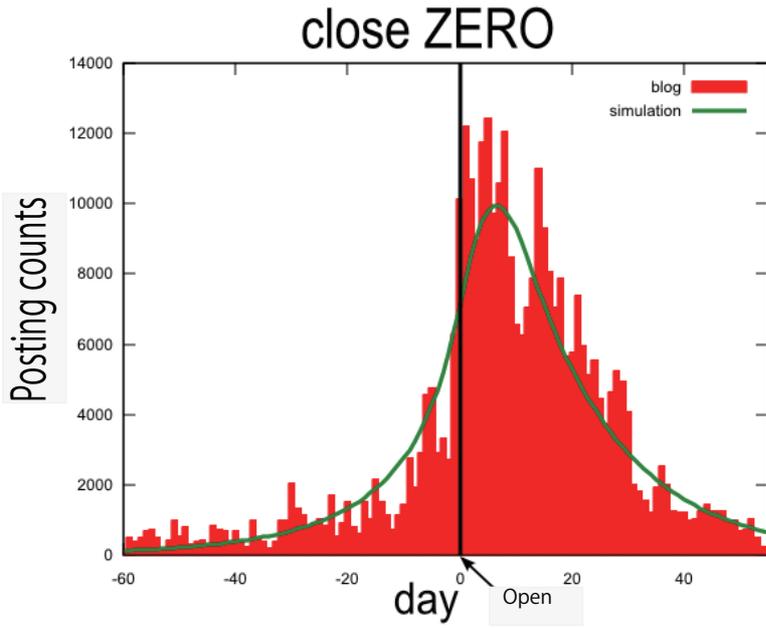

**Figure 13**
Calculation (solid line) and blog posting (histogram) for Close Zero(JPN).
R=0.035

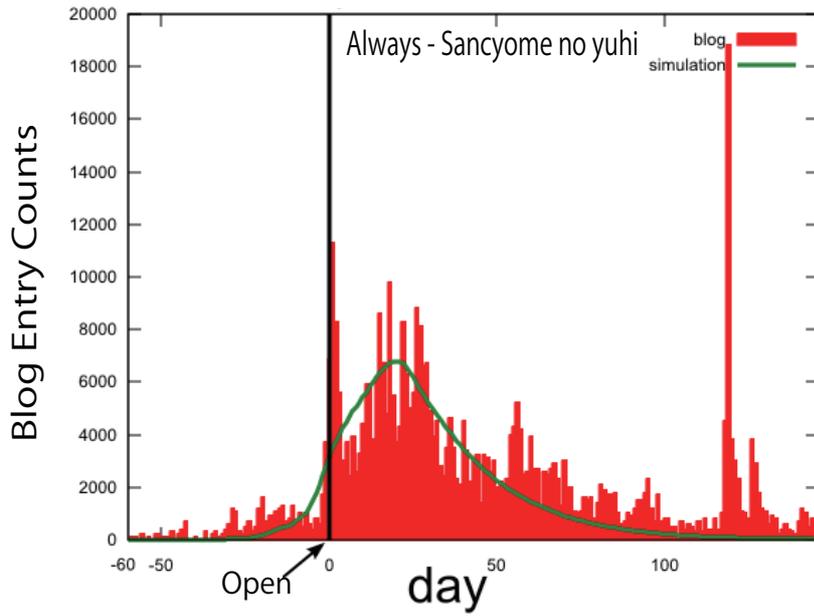

**Figure 14**
Calculation (solid line) and blog posting (histogram) for Always (JPN).
R=0.213

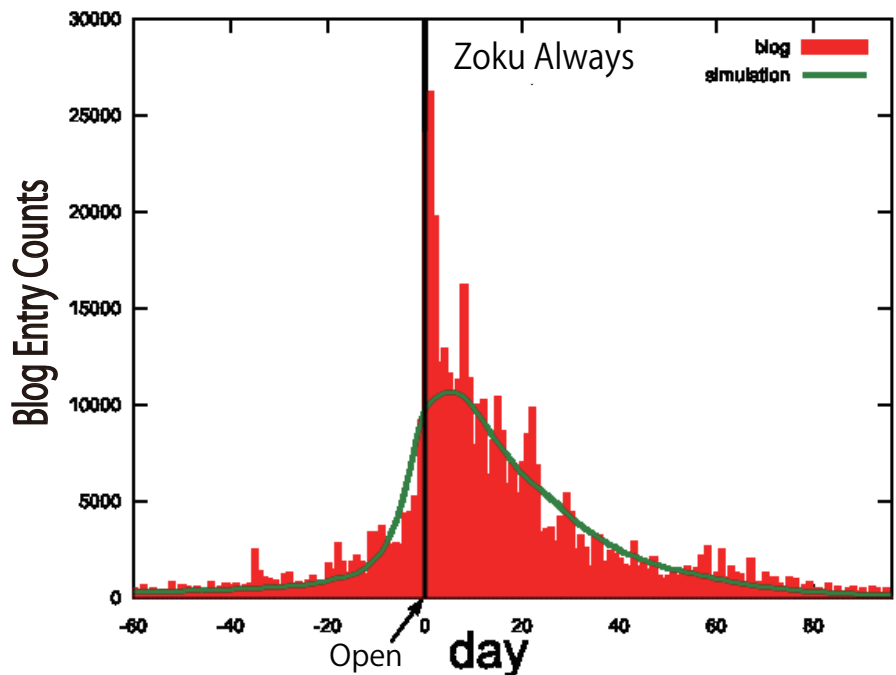

**Figure 15**
Calculation (solid line) and blog posting (histogram) for Zoku-Always (JPN).
R=0.107

## 5. Discussions
*5.1 Agreement with observations*

In figs.11-15, we found that our calculations agree well with the observed daily posting numbers of weblog for each movie. Since we found that the daily posting numbers of weblog for a movie is proportional to the revenue of the movie, this agreement means that our calculation can reproduce the revenue of each movie. For the Japanese market of motion pictures, the new movie should be watched in cinema, because the online download and the DVD/BD disc is available several months after the open day of the movie. Moreover, the entrance fee of cinema in Japanese market is similar price for almost all cinemas. Thus, the agreement of our calculation with the daily weblog posting number means that our calculation can describe the movement of people in the society at least for the Japanese market of the motion picture very well.

The agreement can be considered that our theory can describe the collective motion of people in society at least for the entertainment market in Japan. Since our theory is considered as general, we expect that it can be applied to other entertainment market in the world.

*5.2 Optimization of time distribution of advertisement cost*

As we see in the section 3.1, the advertisement cost has distribution in time. The calculated result using eqs.(28) and (29) of our theory depends strongly on the time distribution of the advertisement cost. It means that, even for the same total cost, the total revenue can be changed depends on the time distribution of the advertisement cost. Thus, using our eqs.(28) and (29), we can consider how to optimize the time distribution of the advertisement cost to get the maximum value of the total revenue at the condition of the fixed total advertisement cost.

*5.3 Indirect communication*

One of the original idea of the present theory is the effect of the indirect communication that is explained in the section 2.3. To demonstrate the effect of the indirect communication, we perform 2 caclulation; calculation including indirect communication and calculation without indirect calculation. The example we show here is the calculation for Avatar which was the very big hit motion picture in 2010. The both calculatons is optimized as much as possible using the Metropolis-like way of the parameter optimization with random numbers to minimize the R-factor. The result is shown in fig.16. The result shows us that the theory is very important to describe human interactions in the real society.

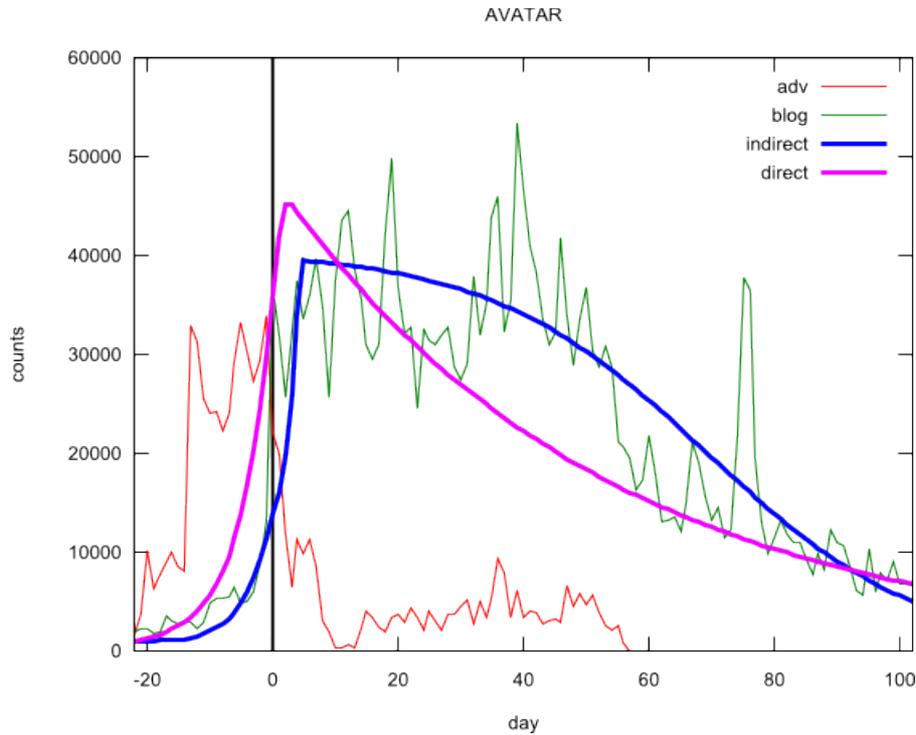

**Figure 16**
The indirect communication effect for Avatar. The day =0 means the open day of the movie Avatar in Japan. "adv" means the daily advertisement cost, "blog" meaans the daily posting number of weblog for the avatar, "direct" means the calculation excluding the indirect communication effect. "indirect" shows the calculation including the indirect communication. For the both calculation, the parameters in the calculation are optimized using the R-factor.

*5.4 Relation to the Bass model*
The Bass model [17, 18] is well-known to be the model of the spread of the WOM. In this subsection, we verify that the Bass model is derived from the equation of the mathematical model of hit phenomena. For the sales of the production, the equation of our mathematical model for hit phenomena is written as follows,

$$\frac{dI_i(t)}{dt} = A_i + \sum_j D_{ij} I_j(t) + \sum \sum P_{ijk} I_j(t) I_k(t) \tag{31}$$

The number of people m and the purchase intention writes the total number of sales for the product N as follows,

$$\sum_i I_i(t) = m \langle I(t) \rangle = N(t) \tag{32}$$

In the Bass model, the advertisement is spread from adaptors to non-adaptors,. Thus,

$$\sum_i A_i = a(m - N(t))$$
$$= \sum_i a \tag{33}$$

where the sum is only for the persons who do not buy the product.

For the direct communication term in eq.(31), $\sum_i \sum_j D_{ij} I_i I_j$, we assume here that the coefficienat $D_{ij}$ is not zero only for the pair of the persons, adaptor and non-adaptor. The purchase intention for the adaptor person is considered to be $I_j = 0$. Thus, we can write the direct communication term as follows,

$$\sum_i \sum_j D_{ij} I_j = (m - N(t)) N(t) d \langle I \rangle$$
$$= (m - N(t)) N(t) b$$

$$\frac{d\sum_i I_i(t)}{dt} = \sum_i \left[ A_i + \sum_j D_{ij}(t) I_j(t) + \sum\sum P_{ijk} I_j(t) I_k(t) \right] \quad (35)$$

Therefore, substituting (34) into (28), we obtain,

$$\frac{dN(t)}{dt} = a(m - N(t)) + (m - N(t))N(t)b$$
$$+ \sum_i \sum_j \sum_k P_{ijk} I_j I_k \quad (36)$$

Here, if we neglect the indirect communication term as $P_{ijk}=0$ we obtain the well-known Bass model equation.

$$\frac{dN(t)}{dt} = a(m - N(t)) + (m - N(t))N(t)b \quad (37)$$

Therefore, we find that our mathematical model for hit phenomena include the Bass model as the case of neglecting indirect communications. Thus, the indirect communication in eq. (21) we included is the new effect for WOM effect.

*5.5 Statistical physics method for human-interactions*

In this paper, we calculate the time variation of the averaged action of human in the real society using our theory. Eq.(21) or eq.(28) and (29) seems to describe well the collective action of human for entertainment. Not only for potion picture business, our theory can be applied to the incoming number of people to local festivals in Japan.[29] As same as the usual physics approach, we calculate the prediction and compare it with observations. Thus, if we obtain a lot of observation data from the real society, the usual method of statistical physics or many-body physics can be available to investigate the human interaction in the real society. Nowadays, it begin to be possible to collect a lot of observations of human action using the social networks like blogs, twitter and Facebook. For this purpose, of course, the human interaction for each individual problems should be investigated carefully with the communication to social scientist.

## 6. Conclusion

We present the mathematical model of hit phenomena as an equation of action of consumer where the consumer-consumer communication is taking into account. In the communication effect, we include both the direct communication and the indirect communication. We found the daily posting number of blog is very similar to the revenue of corresponding movie. The daily number of blog posting can be used as quasi-revenue. The results calculated with the model can predict the revenue of corresponding movie very well. We found that the indirect communication affect the revenue using the calculation using our theory. The conclusion presented in this paper will be applicable to any consumer market.

**Acknowledgement**
Dentsu Inc presents the advertisement cost data. The authors are grateful for Takeshi Nakagawa of Dentsu Inc. for helping us to use the data. The revenue data is presented from the Nikkan Kogyo Tsushinsya. The research is partially supported by the Hit Contents Laboratory Inc. The Okawa Foundation supports the research for Information and Telecommunications.